# Delayed Offloading using Cloud Cooperated Millimeter Wave Gates

[1,2]Ehab Mahmoud Mohamed, [1]Kei Sakaguchi, and [1]Seiichi Sampei
[1]Graduate School of Engineering, Osaka University, [2]Electrical Engineering Dept., Aswan University.
Email: ehab@wireless.comm.eng.osaka.ac-u.ac.jp, {sakaguchi, sampei}@comm.eng.osaka-u.ac.jp

*Abstract*— **Increasing wireless cellular networks capacity is one of the major challenges for the coming years, especially if we consider the annual doubling of mobile user traffic. Towards that and thanks to the fact that a significant amount of mobile data is indeed delay tolerable, in this paper, we suggest embedding the delayed offloading of some user traffic to be a part of future wireless cellular networks. To accomplish this, user delayed files will be offloaded using ultra-high speed Millimeter Wave (Mm-W) Gates. The Mm-W Gate, which will be distributed inside the Macro basestation (BS) area, consists of number of Mm-W Access Points (APs) controlled by a local coordinator installed inside the Gate. To effectively manage the delayed offloading mechanism, utilizing the concept of User/Control (U/C) data splitting, the Gates coordinators and the Macro BS are connected to the Cloud Radio Access Network (C-RAN) through optical fiber links. Also, files offloading organizer software is used by the User Equipment (UE). A novel weighted proportional fairness (WPF) user scheduling algorithm is proposed to maximize the Gate Offloading Efficiency (GOFE) with maintaining long term fairness among the different mobility users pass through the Gate. If the Gate is properly designed and the files delay deadlines are properly set; near 100% GOFE with average reduction of 99.7% in UE energy consumption can be obtained, in time the user just passes through the Gate.**

## I. INTRODUCTION

Due to the huge proliferation of smart phone and tablet users, the cellular network capacity comes at the bottleneck. Many researches claim that at the end of this year, 2014, a broadband mobile user will consume average traffic of 7 GB per month with expected exponential increase of 2-fold per year [1] [2]. Moreover, Cisco expects that by the year 2018, video data acquires 66% of the total mobile traffic [3]. Based upon these future expectations, many researchers pay the attention of how we can increase current cellular network capacity to accommodate the expected huge increase in mobile traffic [1] [2], [4]-[6].

One promising strategy is to offload some mobile traffic through other fixed networks like WiFi networks [2] [5]. Cisco claims that, about 52% of total mobile traffic will be offloaded by the year 2018 and only 48% of the generated traffic will be transmitted through cellular networks [3]. Currently, most of smart phones and tablet users use the on the spot offloading, where the user offloads most of his bulk generated data such as video downloading, mobile backup, etc., whenever he is under WiFi hotspot coverage. Recently, many researchers highlight the concept of delayed offloading based upon the fact that most of mobile traffic is indeed delay tolerable. i.e., the traffic can be delayed to some predefined time without affecting user satisfactions [2] [5] [7]. In [5], using high mobility traces inside 3 different cities, the authors reported that, if we can delay the response to a user traffic request only by 100 sec, to let him enter a WiFi zone, about 20%~30% offloading gain can be obtained over the on the spot offloading. Based upon WiFi availability, UE mobility and files delay tolerances, with 7GB per month user traffic, about 33%~80% of the 3G cellular network traffic can be offloaded through WiFi networks [2][5]. Most of delayed offloading research results, using WiFi networks, come from field measurements, using the deployed WiFi APs in a city, with different mobile users' traces.

Although the studies of delayed offloading using WiFi networks emphasize the high efficiency of delayed offloading mechanism in solving current cellular network capacity problem, the researchers claim that; in order to accommodate the future increase in mobile traffic, cellular network operators should deploy hundred thousands of WiFi APs, which in turn, results in very high congestions and poor offloading performance. In addition, they don't give a complete network framework including the delayed offloading mechanism to the cellular operators, in which the user always generates different file sizes to be uploaded/ downloaded with certain deadlines, and the network always keeps track of this files table update. At the same time, the network sends back information of the nearest offloading zone to the user, and how he can reach it, so the files deadlines can be probably adjusted based upon the next zone, he is able to reach. Then, based upon the user mobility and the delayed files table, the network try to upload /download all the user traffic within the assigned deadlines.

In this paper, we propose the network structure and protocol needed to incorporate the delayed offloading as a part of future cellular networks using ultra-high speed Mm-W Gates. Each Gate consists of number of 60 GHz with 2.16 GHz bandwidth Mm-W APs using IEEE 802.11ad [8] standard controlled by a local coordinator installed inside the Gate [9]. The Gates will be located at the entrance of enterprise, shopping mall, office, building, etc., so they can cover users' daily habits. To effectively manage the delayed offloading mechanism, utilizing the concept of U/C splitting [6], the Gates coordinators and the Macro BS are connected to the C-RAN through optical fiber links. Also, files offloading organizer software is used by the UE.

The use of high speed Mm-W APs enables us to cope with the future increase in mobile traffic; in our simulations, we consider user traffic of 7000 GB per month, the year 2024 expected user traffic. In addition, the user doesn't need to stay much time inside the Gate coverage, in our assumption; we consider that the user just passes through the Gate with his normal speed which normally takes multiple seconds. Therefore, delayed offloading using the Mm-W Gates will greatly reduce the deployment cost over the WiFi solution, because there is no need for the operators to deploy hundred thousands of WiFi APs with high congestion and complicated control. Also, it will greatly save the user time and UE energy consumption, which gets it a promising solution for future cellular network capacity problem. We will give the full network structure and protocol that manage the delayed offloading process using the Gates. Moreover, a novel weighted proportional fairness (WPF) user scheduling algorithm is proposed to increase the Gate offloading efficiency, at the same time, it ensures long term fairness among the different mobility users pass through the Gate.

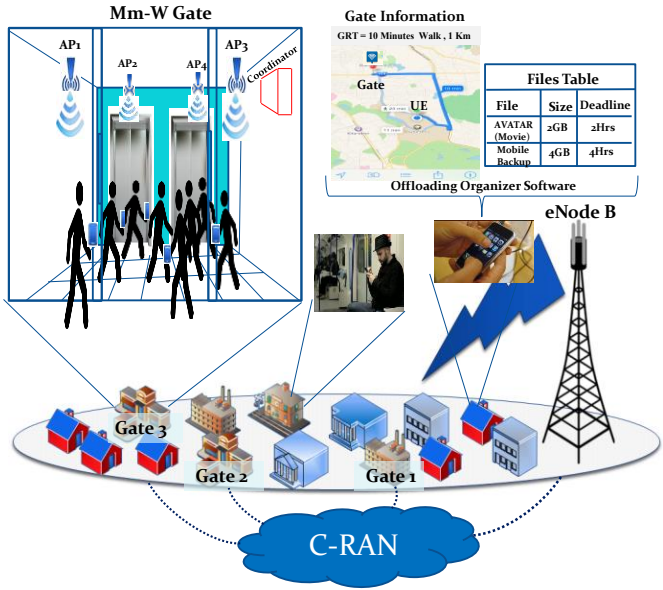
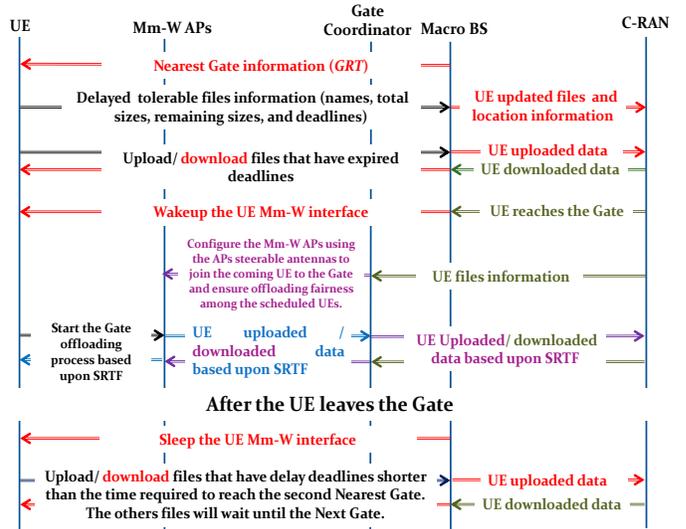

Fig. 1 The network structure including the Mm-W Gates

Fig. 2 The proposed network protocol

The reset of this paper is organized as follows; Section II gives the detailed network architecture and protocol using the Mm-W Gates. User scheduling inside the Gate is given by Section III. Section IV gives the simulation analysis followed by the conclusion in Section V.

## II. THE PROPOSED NETWORK ARCHITECTURE AND PROTOCOL

Figure 1 shows the network structure including the Mm-W Gates. In order to cover usual mobile user daily habits as much as possible, the Mm-W Gates are distributed inside the Macro BS area, and they are located at the entrance of buildings, train stations, airports, enterprise, shopping mall.., etc. The Mm-W Gate (Gate 3), in Fig.1, consists of 4 Mm-W APs controlled by a local coordinator installed inside the Gate [9]. The Gates coordinators and the Macro BS are connected to the C-RAN using optical fiber links. Different mobility users enter the Gate with different delayed file sizes to be uploaded /downloaded in time the user just passes through the Gate. Through using U /C plane splitting [6], UEs delayed traffic will be transmitted using the high speed Mm-W APs, while the control data is transmitted through the large coverage of Macro BS [1] [6].

To effectively handle and control the delayed files transmissions, each UE has an offloading organizer (scheduler) software application that organizes the files offloading process. The software always records the UE generated delayed files, titles, total sizes, remaining delay times (deadlines) and remaining sizes after partial transmission / offloading, in a kind of files table. Based upon the current UE location and its mobility trace, and to ensure load balance among the Gates, the C-RAN finds out the nearest Gate, which lies on the UE path, with capacity and load relevant to the UE delayed traffic size. The C-RAN sends back the information of the nominated Gate to the UE organizer software, through the Macro BS, this information includes the estimated gate reaching time (*GRT*) and how to reach. The user can view the nearest Gate information, *GRT* and how to reach, by using a google map like interface, Fig.1, built in the organizer software. Hence, he can easily go to the nominated Gate within the estimated *GRT*. This operation is adaptively done by the C-RAN based upon the changes of UE mobility trace, its delayed traffic size, and nearest Gates conditions. Usually, the files deadlines are automatically adjusted by the scheduler based upon the UE generated data type and *GRT* of the nominated Gate. That is, if the generated file is of VoiP or Peer to Peer (P2P) data types, zero deadlines will be set by the software, and the transmission is done by the Macro BS. But, for other delay tolerable data types like video, mobile backup, etc., and to let the UE enter the Gate for offloading, the deadlines are adjusted with average greater than *(GRT-FAT)*, where *FAT* is the file arrival time. Sometimes, the user wants to tune the files deadlines based upon his urgencies and desires, to cover such cases, the user can access to the organizer software and assign the files deadlines based upon his needs, so he can manually trigger the scheduler. But, for high UE energy saving, it is better to assign files deadlines greater than *(GRT-FAT)*, especially for the too bulk data type. The delayed files are scheduled for transmission in shortest remaining time first (SRTF). The C-RAN always has the updated version of UEs files tables. Hence, it can advise the UE about the nearest Gate on his path with capacity and load relevant to its currently delayed traffic size. Together, it passes them to the Gate coordinator, whenever some mobile users reach a certain Gate, to be used for data prefetching, users scheduling and files offloading, see Sec III.

The Mm-W APs steerable antennas are used to track the UEs inside the Gate through beamforming. The beamforming mechanism is fully controlled and configured by the Gate local coordinator based upon the proposed user scheduling algorithm.

Figure 2 shows the proposed protocol that organizes the operation between the proposed network basic elements; the UE, the Mm-W APs, the Gate coordinator, the Macro BS and the C-RAN.

## III. THE PROPOSED USER SCHEDULING ALGORITHM

Inside the Gate user scheduling is one of the most challengeable problems in the Gate design because each user has different mobility whenever he passes through the Gate. In consequence, users will have unequal stay time inside the Gate coverage. The

Gate coordinator is responsible of user scheduling. It controls the beam steering of the Mm-W APs to catch the different mobility users based upon the following proposed weighted proportional fairness (WPF) scheduling algorithm.

The Gate user scheduling problem can be formulated as: how we can maximize the Gate offloading Efficiency, with maintaining long term fairness among the unequal stay times' users.

In the definition of fairness, we concern in:

- Allocation Fairness $F_A(\Delta t)$, which considers the amount of allocated resources (time slots) within a given time interval $\Delta t$:

$$F_A(\Delta t) = \frac{(\sum_{k=1}^{K} A_k(\Delta t))^2}{\left(K \sum_{k=1}^{K} (A_k(\Delta t))^2\right)} \quad (1)$$

- Byte Offloading Fairness $F_{BO}(\Delta t)$, which considers the amount of offloaded bytes within a given time interval $\Delta t$:

$$F_{BO}(\Delta t) = \frac{(\sum_{k=1}^{K} BO_k(\Delta t))^2}{\left(K \sum_{k=1}^{K} (BO_k(\Delta t))^2\right)} \quad (2)$$

where $K$ denotes the total number of users inside the Gate, $A_k(\Delta t)$ and $BO_k(\Delta t)$ denote the total number of allocation units (number of time slots) scheduled to user $k$ and the total number of Bytes offloaded by him, during his stay inside the Gate coverage, respectively. $F_A(\Delta t)=1$ and $F_{BO}(\Delta t)=1$, indicate that all users receive equal number of time slots and they offload the same amount of bytes, assuming that all users have infinite number of bytes to be offloaded during their stay inside the Gate coverage.

In order to attain fairness among the different mobility users, we propose a WPF user scheduling algorithm; in which we give more priority to the user who has less time to stay inside the Gate coverage due to his high speed. Hence, the incremental utility function for user $k$ at time slot $n$ can be expressed as [10]:

$$U_k(n) = \frac{r_k(n)}{R_k(n)} [w_k(n)]^\alpha \quad (3)$$

$\alpha \in [0;1]$, $\alpha = 0$, means that the utility corresponds to the standard PF algorithm, but when $\alpha = 1$, means that the utility corresponds to the proposed WPF algorithm. $r_k(n)$ is the instantaneous user $k$ offloading rate at time slot $n$, which can be defined as:

$$r_k(n) = \min(C_k(n), L_k(n)/T_{sl}) \quad (4)$$

where $C_k(n)$ is the capacity (bps) received by user $k$ at time slot $n$, which is directly related to the received Signal to Interference plus Noise Ratio (SINR). $L_k(n)$ is the total remaining size of user $k$ delayed files indexed in his offloading scheduler at time slot $n$, and $T_{sl}$ is the time slot period. The user files table is sent by the C-RAN to the Gate coordinator as the user reaches to enter the Gate, and the C-RAN always keeps track of the table updates. Accordingly, during one time slot period $T_{sl}$, a user $k$ cannot send more data than his available capacity and he cannot send more data than it is indexed in his offloading organizer. By this way, we ensure the maximization of the GOFE as we will see shortly.

$R_k(n)$ is the average offloading rate received by user $k$ up to time slot $n$

$$R_k(n) = \left(1 - \frac{1}{N_c}\right) R_k(n-1) + \frac{1}{N_c} r_k(n) a_k(n) \quad (5)$$

where, $a_k(n) = 1$, if user $k$ is chosen to be scheduled at time slot $n$, and 0 elsewhere, and $N_c$ is the averaging low pass filter response time.

$w_k(n)$ is the user $k$ priority factor at time slot $n$, which is defined as the inverse of user $k$ expected given resources (time slots) from time slot $n$ until it leaves the Gate normalized to the expected longest given resources. The users' assigned resources expectation is directly related to their expected stay times. Hence, $w_k(n)$ can be expressed as:

$$w_k(n) = \left(\frac{TS_k(n)}{TS_h(n)}\right)^{-1} \quad (6)$$

where $TS_k(n)$ is the expected time user $k$ will stay inside the Gate from time slot $n$ until it leaves the Gate, and $TS_h(n)$ is the expected longest stay time belongs to the lowest speed user. The user stay time $TS_k(n)$ can be expected as:

$$TS_k(n) = \frac{d_k(n)}{\overline{v_{k(n)}}} \quad (7)$$

where $d_k(n)$ is the distance between the Gate exit position and user $k$ current position, and $\overline{v_{k(n)}}$ is his average velocity at time $n$.

Therefore, user selection can be done as follows:

For 1-AP

$$\dot{k}(n) = \arg\max_K \{U_k(n)\} \quad (8)$$

In this case, the coordinator evaluates each user utility metric given by Eq. (3), and chooses the user that has the maximum utility value as given by Eq. (8) to be scheduled at the current time slot $n$. User $k$ capacity evaluation is done using beam training [8] or by using UE location estimation [9].

As it is clearly appeared from the proposed scheduling criterion Eq. (8), at time slot $n$, if there are two users with equal expected assigned resources ($w_1(n) = w_2(n)$) and same average offloading rate ($R_1(n) = R_2(n)$), the scheduling priority will be given to the user who currently has higher offloading rate (higher $r(n)$). At the same time, if the two users have the same instantaneous offloading rate ($r_1(n) = r_2(n)$) and same average offloading rate, the priority will be given to the shortest stay time user. At last, if they have the same offloading rate and they have equal expected assigned resources, the priority will be given to the one with lower average offloading rate. Hence, the proposed scheduler always increases the Gate offloading efficiency with maintaining long term fairness among the scheduled users in both resource allocations and byte offloading.

For more than 1-AP

In case of more than one Mm-W AP, the coordinator will jointly choose a subset of users $\dot{M}(n) = [m_1, m_2, ..m_{NAP}]$ to be connected to the assigned APs $NAP$, at time slot $n$, based upon the following equation:

$$\dot{M}(n) = \arg\max_{M \subset K} \left\{\sum_{k=m_1}^{m_{NAP}} U_k(n)\right\} \quad (9)$$

where $M \subset K$ indicates all available user subsets of the users domain $K$. In order to do that, the coordinator constructs a look up table that contains the capacity received at each UE, using all AP-UE available configurations (subsets), through beam training [8] or based upon UEs locations estimation [9], and it chooses the users subset (AP-UE configuration) that maximizes the utility metric given by Eq. (9).

IV. SIMULATION ANALYSIS

In this section, we give the simulations analyses that prove the effectiveness of the proposed delayed offloading network using the Mm-W Gates, in addition to the proposed user scheduling algorithm.

*A. Simulation Scenario and Simulation Area*

In our simulation scenario, we consider one Mm-W Gate with a local coordinator connected and the Macro BS to the C-RAN.

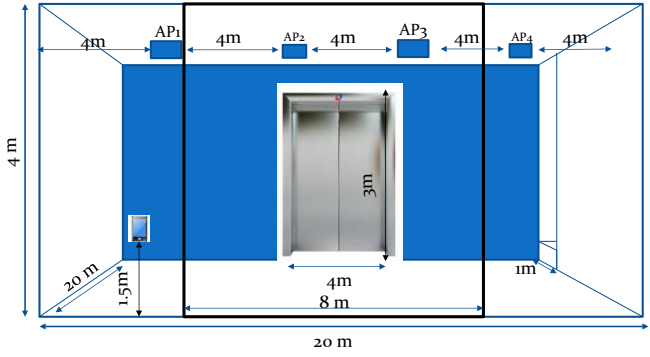

Fig. 3 The Gate ray trace simulation area

We assume that all users reach the Gate with same *GRT* value. Each UE has random number of delayed files to be uploaded/downloaded, indexed in its files offloading organizer software. The files are generated with average traffic density of 1.62 GB/10 minutes (the year 2024 expected mobile traffic). The files deadlines are generated using truncated Gaussian distribution, as it will be explained in the next subsection. The files that have expired deadlines, before the user reaches the Gate, will start to be transmitted using the Macro BS. As soon as, the user reaches the Gate, the ongoing file transmission is switched to be transmitted using the Gate through seamless handover controlled by the C-RAN. The wakeup signal of the UE Mm-W module and others control signals are sent to the UE through the Macro BS. In our simulation assumption, after the user leaves the Gate, the Macro BS will sleep the UE Mm-W module, and all the remaining UE indexed files will be transmitted using the Macro BS. In order to highly emphasize the great impact of the proposed delayed offloading network over the case of no offloading (using Macro BS only), we consider the use of LTE Macro BS with 100 Mbps net transmission speed, which is unrealistic assumption in real scenarios. Figure 3 shows the ray tracing simulation area of the Mm-W Gate. We disperse the Mm-W APs locations to reduce the interference as much as possible especially we assume that all APs use the same channel. Also, to reduce human blocking effect as much as possible, we attached the APs to the ceiling. In the ray tracing simulation, all Gate materials are from concrete except the Gate entrance door and the elevator are from glass and metal.

In our analyses, we concern on measuring the GOFE using different number of Mm-W APs and different *GRT* values, which can be defined as:

$$GOFE = \frac{\sum Bytes\ Successfully\ Transfered\ by\ the\ Gate}{Total\ UEs\ Geneared\ Bytes} \quad (10)$$

Also, we concern in measuring the average UE energy consumption (energy consumption due to Macro BS usage + energy consumption due to Gate usage) normalized to the case of no offloading (using Macro BS only). Also, the effectiveness of the proposed WPF in comparison to the conventional PF and Round Robin will be investigated.

*B. The Files Deadlines Distribution Model*

As previously explained, the organizer software automatically adjusts the files deadlines based upon the generated file type, unless the user himself assigns some files deadlines based upon his desires. Based upon this scenario, we can roughly model the files deadlines generation as a truncated Gaussian random process (truncated at zero deadlines). The mean of the Gaussian distribution equals $\rho(GRT - FAT)$ and standard deviation of $\delta$. $\rho$ and $\delta$ are different from user to user based upon user's needs.

Table .1 The simulation parameters

| Parameter | Setting |
|---|---|
| Num. of Mm-W APs | 1,2,3 and 4 |
| Num. of UE inside the Gate | 14 |
| AP bandwidth | 2.16 GHz |
| AP Tx Power | 10 dBm |
| User traffic model | Exponentially random file sizes of average 1.62 GB with exponentially random interarrival time (IAT) of 10 minutes average (traffic density by the year 2024) |
| UE mobility model inside the Gate | Random Walk with average speed of 5Km/Hr. |
| Bandwidth and SNR efficiencies [11] | 0.7 and 1 |
| Time slot period | 3 m sec |
| Files deadline model | Gaussian distribution truncated at zero. |
| Gate Reaching Time (*GRT*) | 0.5, 1, 1.5 and 2 Hrs |
| Macro BS transmission rate | 100 Mbps |
| Macro BS Tx power | 46 dBm |
| Probability of path blocking due human motion [8] | Uniform distribution |
| $\rho, \delta$ | 1.5, 0.1(*GRT-FAT*) |

For high Gate offloading efficiency and UE energy saving, with coinciding with Cisco report of percentages of user generated files' types [3], more than 90 % of files deadlines should be greater than (*GRT-FAT*) value. Hence, the Gaussian distribution parameters should be $\rho > 1$ and $\delta < (\frac{\rho-1}{2})(GRT-FAT)$. In our simulations, we set $\rho = 1.5$ and $\delta = 0.1(GRT - FAT)$, that ensure more than 90 % of users' generated files are reaching the Gate before deadlines expiration, other simulation parameters are in Table. 1.

*C. Simulation Results*

Figure 4 shows the GOFE using different number of Mm-W APs and different *GRT* values of 0.5Hr, 1Hr, 1.5Hr and 2Hrs. From Fig. 4, as the number of Mm-W APs is increased, the GOFE is increased. Also, as the *GRT* is increased the GOFE is decreased using the same number of APs. This comes from the huge number of delayed files each user generates as the *GRT* increased. By considering that the Gate offloading is only done during the Gate passing time, many of the users' generated traffic will be transmitted through the Macro BS which will highly reduce the GOFE. From Fig. 4, using 4 APs with 0.5 Hr *GRT* (almost 68 GB total users generated traffic), near 100% GOFE can be obtained. Actually the GOFE is slightly less than 100 % due to the zero deadlines files in the truncated Gaussian model. Figure 5 shows UE average energy consumption using delayed offloading with Mm-W Gate normalized to the case of no offloading (using Macro BS only) with different number of APs and different *GRT* values.

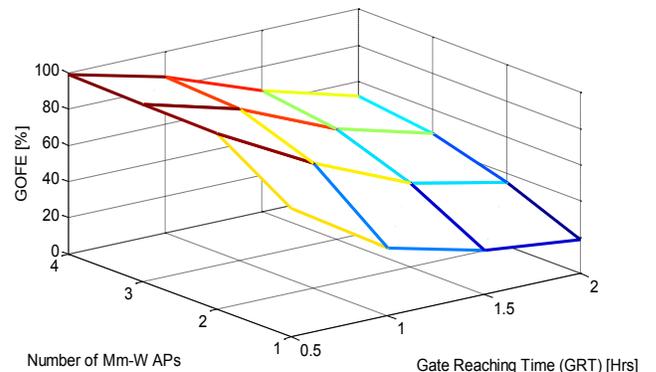

Fig.4 Gate offloading efficiency (GOFE)

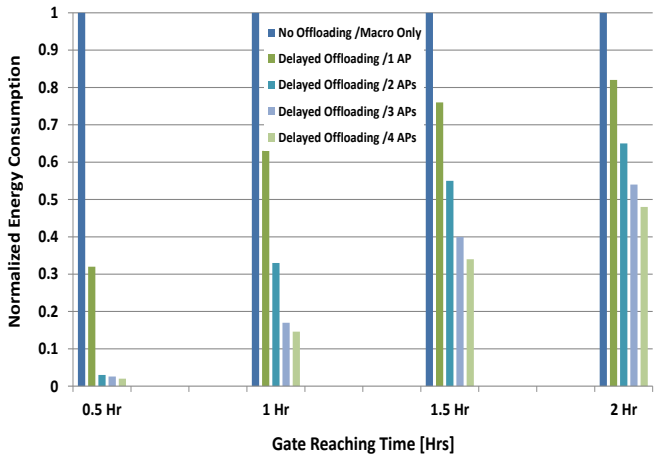

Fig.5 UE normalized average energy consumption

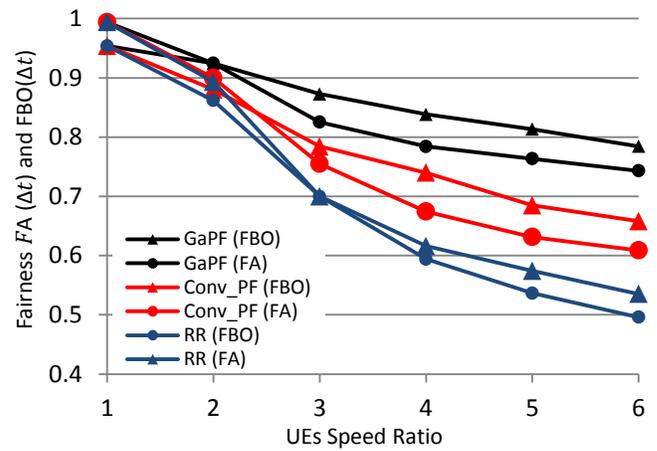

Fig.6 Allocation and Byte offloading fairness comparisons

The normalized energy consumption can calculated as:

$$\frac{\text{Energy of delayed offloading (Gate Usage + Macro BS Usage)}}{\text{Energy of No offloading (Macro Only)}}$$

The energy consumption is inversely proportional to the GOFE. As the GOFE is increased, the normalized energy consumption is highly decreased and vice versa. It is interesting to find that average normalized energy consumption of 0.03 can be reached in case of 0.5 Hr *GRT* and 4 APs in Gate coverage time of multiple seconds. These results confirm the high impact of the Mm-W Gats as a solution for future cellular networks capacity problem.

Figures 6 and 7 confirm the effectiveness of the proposed Gate WPF (GaWPF) user scheduling algorithm over the conventional PF (Conv_PF) and round robin (R.R). The scheduling Algorithms are tested over the total time span of Gate coverage. In addition, the tests are done using different UEs speed ratios, where speed ratio is the ratio between the highest speed UE and the lowest speed UE, which is direct reflection of the users unequal stay time, if all users move in one direction towards the Gate exit. Also, 3 APs Gate with 1 Hr *GRT* are assumed. From these figures; always the proposed WPF has the best performance among the other candidates in Allocation Fairness, Byte offloading Fairness and GOFE. This is because the proposed scheme takes into account the users' mobility, when it jointly chooses the subset users for current scheduling Eq. (9).

## V. CONCLUSION

In this paper, as a contribution to solve the future cellular networks capacity problem, we proposed to embed the delayed offloading mechanism to be a part of future cellular networks. Towards that, we proposed to use Mm-W Gates inside the Macro BS area. We gave the detailed architecture of the proposed network in addition to the protocol that organizes the operation between its elements. Then, we highlighted the user scheduling problem inside the Gate. An efficient WPF user scheduling algorithm is proposed to solve this problem. We proved the high impact of the proposed network in terms of GOFE and average UEs energy consumption. Also, we proved the high efficiency of the proposed WPF user scheduling over the conventional PF and round robin user scheduling algorithms.

ACKNOWLEDGMENT

This work is partly supported by "The research and development project for expansion of radio spectrum resources" of MIC, Japan with project number {0155-0156}, and is also supported by a project named "Millimeter-Wave Evolution for Backhaul and Access (MiWEBA)" under international cooperation program of ICT-2013 EU-Japan funded by FP7 in EU and MIC in Japan.

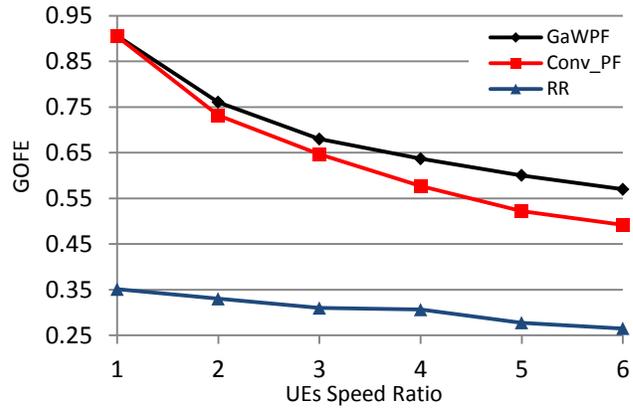

Fig.7 GOFE comparisons